\documentclass[acus]{JAC2003}

\usepackage{graphicx}
\usepackage{booktabs}
\usepackage{verbatim}
\usepackage{subcaption}
\begin{document}

\title{Design Studies with DEMIRCI for SPP RFQ}

\author{B. Yasatekin\thanks{byasatekin@ankara.edu.tr}, G. Turemen, Ankara University, Ankara, Turkey\\  
A. Alacakir, TAEK-SANAEM, Ankara, Turkey\\ 
G. Unel, UCI, Irvine, California, USA\\
}

\maketitle

\begin{abstract}
To design a Radio Frequency Quadrupole (RFQ) is a onerous job which requires a good understanding of all the main parameters and the relevant calculations. Up to the present there are only a few software packages performing this task in a reliable way. These legacy software, though proven in time, could benefit from the modern software development tools like Object Oriented (OO) programming. In this note, a new RFQ design software, DEMIRCI is introduced. It is written entirely from scratch using C$^{++}$ and based on CERN's OO ROOT library. It has a friendly graphical user interface and also a command line interface for batch calculations. It can also interact by file exchange with similar software in the field. After presenting the generic properties of DEMIRCI, its compatibility with similar software packages is discussed based on the results from the reference design parameters of SPP (SNRTC Project Prometheus), a demonstration accelerator at Ankara, Turkey. 
\end{abstract}

\section{Introduction}
A new project in the form of a computer code, written in C$^{++}$,
called DEMIRCI%
\footnote{meaning ``blacksmith'' in Turkish%
} \cite{demirci} is started to explore the potential of the modern
concepts such as object oriented programming and ROOT environment
\cite{root}. This tool helps the designer to create an RFQ model
which would achieve certain goals such as a final target energy or
a fixed total accelerator length in a fully graphical environment.
It calculates a large number of design and beam dynamics parameters
such as energy at the end of the cavity, power dissipation and cavity
quality factor for each cell. It also allows the designer to visualize
a large set of parameters change along the RFQ. Another property of
this tool is the interoperability with similar software in the field,
either directly using the user interface or by simple exchange of
plain text files. 

\section{Design Procedure}

The classical procedure used in designing 4-vane RFQs has been around since LANL designed the first proof of principle (PoP) device. This procedure is known as the ``LANL Four Section Procedure (FSP)'' method \cite{fsp}. According to this method, the RFQ is divided into 4 sections named as radial matching section (RMS), shaper section, gentle buncher section and acceleration section.

\textit{\emph{The potential between the electrodes of a single RFQ
cell is given}}\cite{Biscari}\textit{\emph{ by}}

\begin{eqnarray}
U(r,\theta,z) & = & \frac{V}{2}\Bigg[\sum_{m=1}^{\infty}A_{0m}(\frac{r}{r_{0}})^{2m}\cos(2m\theta)\label{eq:generic-potential}\\
 & + & \sum_{m=0}^{\infty}\sum_{n=1}^{\infty}A_{nm}I_{2m}(nkr)\cos(2m\theta)\cos(nkz)\Bigg]\nonumber 
\end{eqnarray}
where $r$ and $\theta$ are spherical coordinates for which $z$ represents the beam direction, $V$ is the inter-vane voltage, $k$ is the wave parameter given by $k\equiv2\pi/\lambda\beta$, with $\lambda$ being the RF wavelength and $\beta$ being the speed of the ion relative to the speed of light. Also, $r_{0}$ is mean aperture of the vanes, $I_{2m}$ is the modified Bessel function of order 2m and the $A_{nm}$ are the multipole coefficients whose values depend on the vane geometry.

\subsection{New Design Procedure}

The parameters needed to define an RFQ can be divided into two categories: the ones which can be a function of RFQ length and the ones which are constant for a given RFQ. The resonant frequency ($f$), the initial ion energy ($E_{in}$), the input beam current ($I$) and the braveness factor (in terms of the Kilpatrick value) can be cited as examples to the latter. The four parameter vectors falling into the first category are: the synchronous phase ($\phi$), the cell modulation ($m$), the minimum bore radius ($a$) and the inter-vane voltage ($V$). This last one, together with $R/\rho$ could be kept constant along the RFQ length to simplify the design and manufacture.

\begin{figure}[!htb]
\centering
\includegraphics[width=8cm]{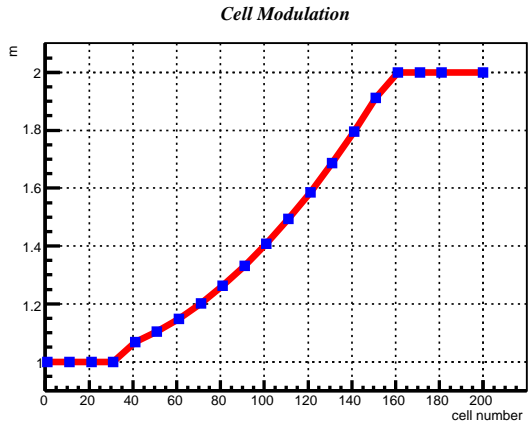}
\caption{Cell modulation ($m$) versus cell number. $m$ is a typical RFQ parameter which changes with cell number. The meaning of blue squares and red line is explained in the text.}
   \label{fig:Cell-modulation}
\end{figure}

 In case of DEMIRCI, a typical parameter's variation along the RFQ can be seen in Fig. \ref{fig:Cell-modulation}. The points represented by the blue squares in Fig. \ref{fig:Cell-modulation} are the so called ``reference cells'' for which the values of the four key parameters are defined by the designer. In this particular example, Fig. \ref{fig:Cell-modulation} shows 20 reference cells for an RFQ of 200 cells in total. The values of the parameters at the cells in between the reference ones are obtained by interpolation assuming a simple linear function. The number of reference cells and the total number of RFQ cells are all user defined variables. The designer might choose to define the values of the parameters for each cell, or to simply define boundary conditions for different regions of the RFQ and let the interpolation function do the rest. 

As a safety check, the software library ensures the monotonic increase of the reference cell numbers. Therefore a new design can be made in a pure graphical way, by simply relocating individual reference cells by using the mouse pointer. This new paradigm allows quick testing of various design ideas concerning the four critical parameter vectors: $\phi$, $m$, $a$ and $V$. Although the non-graphical user interface, i.e. the command line, also exists and it could be more adequate for parameter scan studies, the graphical method has proven itself to be both more intuitive and more pedagogical for the new designers.

\begin{figure}[!htb]
\centering
\includegraphics[width=8cm,height=5cm]{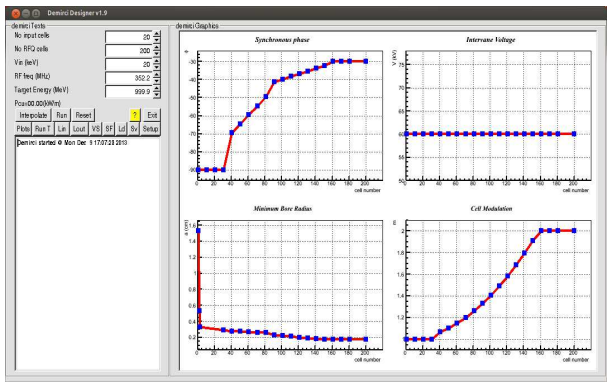}
\caption{The graphical interface of DEMIRCI.}
\label{fig:demirwindow}
\end{figure}

Remaining parameters can be optimized by specifying a goal such as a desired
output beam energy. The number of RFQ cells can be changed from the
default value of 200 to allow the design of longer RFQ cavities. A
shorter design is simpler to obtain by setting a target value for
the exit ion energy. All the scalar RFQ parameters can be tuned using
the number entry boxes at the upper left side of the designer window
which can be seen in Fig. \ref{fig:demirwindow}.

\begin{figure}[!htb]
\centering
\includegraphics[width=8cm,height=5cm]{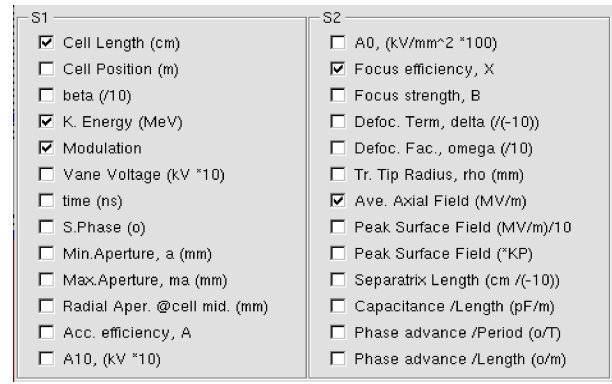}
\caption{A screenshot from DEMIRCI: Drawable parameter selection.}
\label{fig:Plottables}
\end{figure}

Additionally, there are checks along the calculations which inform
the user that a critical parameter is overrun. For example, the designer
is warned if the inter-vane voltage is too high and induces an electric
field above the preselected Kilpatrick limit. If such a warning is
given, it is up to the designer to check the details of the problem
by plotting the parameter in question and to solve it by re-optimizing
the RFQ. Building the RFQ designer software on top of the pre-existing
ROOT libraries provides all the non-essential but necessary functionality,
such as defining the parameters of interest, loading and saving of
the configuration and output files. Additionally, all the graphics
routines in DEMIRCI are based on the ROOT libraries. This design decision
allows a robust, mature and multi-platform GUI experience for the
designer. A section of the user interface dealing with parameter selection
is shown in Fig. \ref{fig:Plottables}. DEMIRCI provides easy plotting of the evolution of the relevant parameters along the RFQ. The graphical results can also be easily customized (such as the formatting of the curves or the addition of a gridline) according to the taste of the user. 



\begin{figure}[!htb]
\centering
\includegraphics[width=5cm]{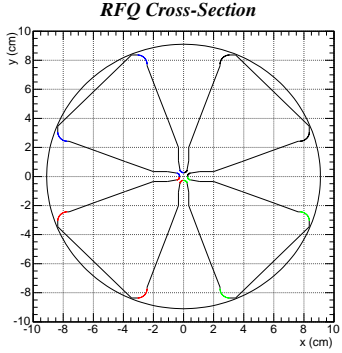}
\caption{The cross sectional view of the RFQ design.}
\label{fig:sfish}
\end{figure}

\begin{figure}[!htb]
\centering
\includegraphics[width=7cm]{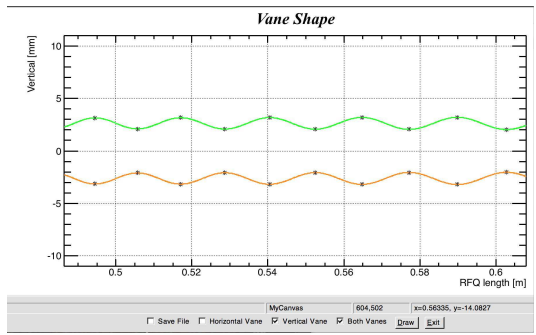}
\caption{The RFQ vane shapes in the y-z plane.}
\label{fig:vanes}
\end{figure}

On the other side DEMIRCI is compatible with TOUTATIS \cite{toutatis} and LIDOS \cite{Lidos}. The files which are built with DEMIRCI can be run on these software packages. Another  property of DEMIRCI is to generate input files for Poisson SUPERFISH 2D simulations \cite{sfish}. The 2D cross section of the designed  RFQ can also be plotted for visual inspection. Such a plot can be seen in Fig. \ref{fig:sfish}. Lastly, it can produce the horizontal and vertical vane shapes which can be fed into 3D solvers for more accurate electromagnetic and thermal studies based on finite element analysis techniques. 
Fig. \ref{fig:vanes} contains such a  drawing focussed on few vertical cells. The points marked by a star are the vane tip extrema and the continuity of the line is achieved by individual sine functions for each cell.
DEMIRCI has the possibility for yielding the coordinates of any point on the vane tip curve and if needed saving the coordinates in a file with a 10 micron accuracy.

\subsection{SPP RFQ as a Design Example}

The SPP RFQ, at TAEK's SANAEM, aims to gain the necessary knowledge and experience to construct a proton beamline needed for educational purposes. A Proof of Principle (PoP) accelerator with modest requirements of achieving at least 1.5 MeV proton energy, with an average beam current of at least 1 $mA$ is under development. This PoP project has also the challenging goal of having the design and construction of the entire setup in Turkey within three years: from its ion source up to the final diagnostic station, including its RF power supply. There are also two secondary goals of this project: 1) Training accelerator physicists and RF engineers on the job; 2) To encourage local industry in accelerator component construction. The design requirements of this machine can be found in Table \ref{tab:SANAEM-Pop}. The input energy was selected to keep the RFQ short for a 1.5 MeV output energy
and at the same time to satisfy the current requirements. The operating frequency was selected to be compatible with similar machines in Europe and therefore to benefit from the already available RF power supply market. Other parameters such as the inter-vane voltage, Kilpatrick value etc. were chosen to be adequate for a first time machine.

\begin{table}[!htb]
\caption{SPP RFQ design parameters.\label{tab:SANAEM-Pop}}
\centering{}%
\begin{tabular}{c|c}
\textbf{Parameter} & \textbf{Value}\tabularnewline
\hline 
\hline 
$E_{in}$(keV) & 20\tabularnewline
\hline 
$E_{out}$ (MeV) & 1.5\tabularnewline
\hline 
$f$ (MHz) & 352.2\tabularnewline
\hline 
$V$ (kV) & 60\tabularnewline
\hline 
$I$ (mA) & >1\tabularnewline
\hline 
KP & 1.5\tabularnewline
\hline 
$R_{0}$ (mm) & 2.8\tabularnewline
\hline 
$\rho$ (mm) & 2.5\tabularnewline
\end{tabular}
\end{table}

\begin{figure}[!htb]
\centering
\includegraphics[width=1\columnwidth]{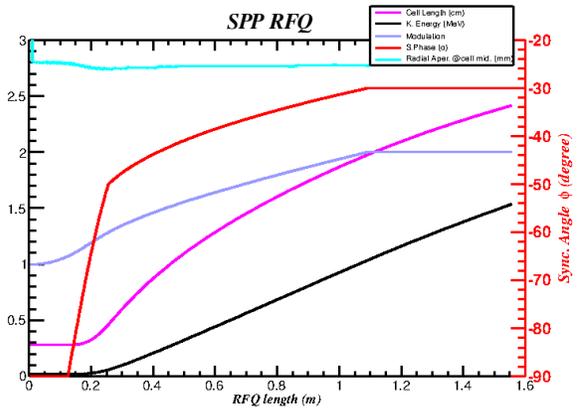}
\caption{SPP RFQ beam dynamics parameters as calculated by DEMIRCI.\label{fig:SANAEM-POP-RFQ}}
\end{figure}

The calculation and visualization of the evolution of the most critical accelerator cavity and beam parameters along the RFQ length was the SPP RFQ which is DEMIRCI's one of the very first applications. The plot containing the evolution of the cell length, beam energy, modulation, synchronous phase and radial aperture can be found in Fig. \ref{fig:SANAEM-POP-RFQ}.

\section{Conclusions}

DEMIRCI is a fast, Unix based modern tool using graphical techniques for RFQ design. It uses analytical formulae based on two term potential to compute the light ion beam behaviour in an RFQ. It also permits
the user to achieve optimizations with specific goals such as a final accelerator length or a final ion beam energy. It interacts with similar software in the field, for result cross check and for further study of the RFQ and beam properties. 

A number of additions and enhancements are being planned for this new tool. The first goal is to use the more complex 8 term potential to allow a more realistic calculation of the EM fields inside the RFQ. This enhancement is expected to further reduce the small deviations in the results obtained with this tool and similar ones. Furthermore, addition of beam dynamics calculations would make DEMIRCI a more complete solution for the RFQ design.

\section{Acknowledgements}

This project is supported by TUBITAK under grant ID 114F106. The authors are also grateful to TAEK for their endorsements and to LANL for their encouragements.


\begin{thebibliography}{9}   


\bibitem{demirci}G. Turemen, G. Unel, H. B. Yasatekin, DEMIRCI: An RFQ Design Code Description, In Preparation.

\bibitem{root}R. Brun and F. Rademakers, ROOT - An Object Oriented Data Analysis Framework, Proceedings AIHENP'96 Workshop, Lausanne, Sep. 1996, Nucl. Inst. \& Meth. in Phys. Res. A 389 81-86, 1997.

\bibitem{fsp}K.R. Crandall, LANL Internal Report, Nr. LA-UR- 96-1836, Revised December 7, 2005. 

\bibitem{Biscari}C. Biscari, Computer Programs and Methods for the Design of High Intensity RFQs, CERN/PS 85-67 (L1), CERN, 1985. 

\bibitem{toutatis}R. Duperrier, ``TOUTATIS: A Radio Frequency Quadrupole
Code'', Phys. Rev. Vol.3, 124201, 2000.

\bibitem{Lidos}LIDOS.RFQ.DESIGNER Version 1.3, http://www.ghga.com/accelsoft/

\bibitem{sfish} K. Halbach and R. F. Holsinger, SUPERFISH - A Computer Program for Evaluation of RF Cavities with Cylindrical Symmetry, Particle Accelerators 7, 1976. 


\end{thebibliography}
\end{document}